\documentclass[preprint,12pt]{elsarticle}

\usepackage{amssymb}

\usepackage[utf8]{inputenc}
\usepackage{hyperref}
\usepackage{multicol}
\usepackage[inline]{enumitem}
\usepackage{amsmath}
\usepackage{amsfonts}
\usepackage{graphicx}
\usepackage{algpseudocode}
\usepackage{algorithm}
\usepackage{gensymb}

\journal{Engineering Applications of Artificial Intelligence}

\begin{document}

\begin{frontmatter}


\author[inst1]{Zacharaya Shabka\corref{cor1}}
\ead{uceezs0@ucl.ac.uk}
\cortext[cor1]{Corresponding Author}

\title{One-shot, Offline and Production-Scalable PID Optimisation with Deep Reinforcement Learning}


\author[inst2]{Michael Enrico}
\author[inst2]{Nick Parsons}
\author[inst1]{Georgios Zervas}

\affiliation[inst1]{organization={University College London},
            addressline={Roberts Building, Torrington Place}, 
            city={London},
            postcode={WC1E 7JE}, 
            country={United Kingdom}}

\affiliation[inst2]{organization={Huber+Suhner Polatis},
            addressline={332 Cambridge Science Park, Milton Road}, 
            city={Cambridge},
            postcode={CB4 0WN}, 
            country={United Kingdom}}

\begin{abstract}
Proportional-integral-derivative (PID) control underlies more than 97\% of automated industrial processes. Controlling these processes effectively with respect to some specified set of performance goals requires finding an optimal set of PID parameters to moderate the PID loop. Tuning these parameters is a long and exhaustive process. A method (patent pending) based on deep reinforcement learning is presented that learns a relationship between generic system properties (e.g. resonance frequency), a multi-objective performance goal and optimal PID parameter values. Performance is demonstrated in the context of a real optical switching product of the foremost manufacturer of such devices globally. Switching is handled by piezoelectric actuators where switching time and optical loss are derived from the speed and stability of actuator-control processes respectively. The method achieves a $5\times$ improvement in the number of actuators that fall within the most challenging target switching speed, $\geq 20\%$ improvement in mean switching speed at the same optical loss and $\geq 75\%$ reduction in performance inconsistency when temperature varies between $5\degree C$ and $73\degree C$. Furthermore, once trained (which takes $\mathcal{O}(hours)$, the model generates actuator-unique PID parameters in a one-shot inference process that takes $\mathcal{O}(ms)$ in comparison to up to $\mathcal{O}(week)$ required for conventional tuning methods, therefore accomplishing these performance improvements whilst achieving up to a $10^6\times$ speed-up. After training, the method can be applied entirely offline, incurring effectively zero optimisation-overhead in production.

\end{abstract}




\end{frontmatter}


\section{Introduction}
\label{section:intro}
Proportional-integral-derivative (PID) control remains one of the most widely used and reliable means of implementing online system control. It is used extensively in many industries from oil refinement to paper production and accounts for approximately 97\% of control processes in industry \cite{techbriefs_PID, Honeywell2002}. PID has advantages with respect to both optimisation/tuning (it only has 3 parameters to optimise) and computation (each control iteration only involves a few simple operations which can easily be implemented on low-cost/high-frequency hardware such as FPGAs or ASICs). However, the best means of optimally tuning PID parameters still undermines it's application. In general, tuning faces a compromise between long and exhaustive but highly optimal methods vs. fast and efficient but non-optimal ones. 

Furthermore, in real world commercial scenarios, products/devices using control loops are manufactured at scale, where the performance of the product/devices can be at least partially if not dominantly dependent on the performance of the closed-loop control process used. For example, the work presented here is done in the context of piezoelectric-actuator based optical switching devices. Current models of switches can be built with up to 768 actuators (384 ports on both the input and output plane), where each is controlled by 2 distinct PID loops (1 per axis). This means that to optimally control each actuator in a single switch of this size, 1536 distinct sets of PID parameters need to be determined. These switches benefit from fast and stable reconfiguration times and low optical loss. Both of these properties depend strongly on the control process underlying these switching processes.

Optimising PID parameters for a large number of non-identical devices faces three primary difficulties. Firstly, since no manufacturing process is perfect, no two manufactured devices will be identical. The subtle but definite differences can (as will be seen in this work) have significant impact on how well they can be controlled by the same set of PID parameters, motivating a means of having unique PID parameters for each device rather than a single generic set. An efficient optimisation method would be able to exploit this device-level information, and use it effectively to generate parameters that are suitable for that device.

Secondly, since the number of devices manufactured can be arbitrarily large, it is desirable to minimise the amount of time it takes to generate these parameters to avoid significant production overhead due to optimisation. Devices can potentially undergo a large number of possible control processes in their lifetime. For example, in a $384\times384$ all-to-all switch where each actuator in each plane can move from pointing towards any position in the opposite plane, to any of the remaining 383 positions, each actuator has 147,456 possible movements it can make per axis - almost 300,000 total per actuator. Since each actuator switches at the order of $\mathcal{O}(10ms)$, checking each of these movements for a single set of PID parameters will take at least 50 minutes. When a large number of combinations in a search process is being explored, it is clear to see how this can easily incur days of overhead. An ideal optimisation routine would not require explicit exposure to each of these movements in order to evaluate if a set of parameters are suitable.

Thirdly, dynamic PID loops, where PID parameters are constantly adjusted over the lifetime of a control process based on the closed loop response of the system, are not suitable in the case of low-cost/high-speed electronics like FPGAs, since they incur additional in-loop computation requirements in order to re-calculate PID parameters. As such it is desirable to find single set of parameters per-device that achieves good control outcomes over it's lifetime and with respect to potentially multiple different performance metrics.

Direct-search based methods have dominated tuning methodologies for many decades \cite{ziegler1942optimum, cohen1953theoretical}. However, these methods must be repeated each time a set of optimal parameters is to be found (i.e. for a new device) and often require long monitoring cycle  to iterate over various parameter combinations to evaluate performance in comparison to some set criteria. They are also difficult to implement in the context of multiple simultaneous (and possibly contending) performance goals since they are typically designed for a particular control outcome.

To summarise - the ideal method of PID parameter optimisation should be able to generate PID parameters such that:
\begin{enumerate*}
    \item each device has a unique set of parameters that are optimal for that device specifically;
    \item these parameters can be generated in a timely manner, not requiring long tuning times or extensive closed-loop operation of the device to do so;
    \item these parameters are determined once for each device with respect to a flexible and multi-faceted performance requirement and are consistent in the face of environmental and operational variability.
\end{enumerate*}

This paper presents a method (patent pending) based on deep reinforcement learning (DRL) that implements one-step, offline and instantaneous optimisation of PID parameters. The method is trained ($\mathcal{O}(hours)$) on a set of devices where it learns a relationship between device information (e.g. resonance-per-axis etc), a multi-objective performance criteria and PID parameter values. After training, the method can be applied to previously unseen devices in a one-shot and offline inference procedure ($\mathcal{O}(ms)$) where some previously measured information (e.g. during post-manufacturing characterisation processes) about the device can be used to directly generate PID parameters that are performant for that device specifically. In this way the method incurs effectively zero optimisation overhead as optimisation time for large numbers of devices is trivial and can be done in parallel to some other process once device information has been measured.

Compared to a direct-search based tuning method implemented in the production setting of a world leading optical switch manufacturer, our method ensures that $5\times$ more switching events are equal to or less than the most challenging target switching time whilst improving average switching time by 23\%. The standard deviation of switching times also improves by 45\%, allowing for more consistent switching performance as well as better performance on average. Moreover, the method is also able to achieve $3.5\times$ greater thermal stability across temperatures ranging from $5\degree$C to $73\degree$C. In addition to this, the proposed method takes $\mathcal{O}(hour)$ to train and $\leq \mathcal{O}(ms)$ to generate new unique parameters for previously unseen actuators. By contrast,  manual (direct search) tuning takes $\mathcal{O}(week)$ to calculate a single set of control parameters for a given actuator and must be re-run if it is to be used on a per-device basis; otherwise using generic parameters leads to (as seen in section \ref{section:results}) much more inconsistent performance. The proposed method is able to achieve a $10^6\times$ speed up when generating device-specific PID parameters that achieve better all around multi-objective control-performance.
\section{Related Work}

Classical PID tuning methods have historically been based on a cost-function driven search process \cite{ziegler1942optimum, zhuang1993automatic, wang1999pid}. These processes are capable of producing high quality parameters. However, such methods often rely on having reliable system models, which is often not possible. They are also slow, requiring a large number of iterations before they find good parameters. This may be acceptable for one-off optimisation processes, but it is prohibitively slow and costly when large numbers of systems have to be individually optimised, where longer production-time per-system incurs additional cost. Finally, these methods are often designed to handle only single performance objectives and their application becomes more complex when multiple, possibly conflicting, objectives are to be simultaneously handled.

More recent optimisation techniques that can handle multi-objective criteria automatically without requiring exhaustive search have been presented as promising PID auto-tuning techniques. Evolutionary/swarm optimisation techniques such as particle swarm or genetic algorithms have been applied to various formulations of the PID tuning problem \cite{aranza2016tunning, Asifa2010, Uren2012}, since they are computationally more efficient than direct search, have good convergence properties and can handle multi-objective optimisation criteria flexibly. However, one fundamental issue with such meta-heuristic algorithms is that the full optimisation process has to be implemented every time a set of parameters is to be found, meaning it is not appropriate when minimising optimisation time is desirable and a large number of distinct devices/systems are to be optimised.

Another general short-coming that applies to all of the methods mentioned above, is the lack of generlisability in the optimisation process. Consider the case when PID parameters need to be found for many devices which are similar (e.g. the devices are the same model of device but are distinguished by inevitable manufacturing imperfections). In this case, it can be reasonably expected that PID parameters would be similar, and that sufficient exposure to a large number of such devices should be able to be exploited in order to more efficiently find parameters for such devices. This premise is not accounted for in the above methods, wich instead must be re-run for each application.

DRL has emerged as another promising means of system control. It has been demonstrated to be able to learn very complex control/operational policies that can yield superior results compared to top human performers in considerably complex and uncertain environments \cite{Mnih2013,Silver2017}. While DRL can in principle be used to control a system directly, replacing PID loops altogether, the decision making process involves a forward pass through a neural network - a process that generally requires expensive and power-hungry hardware such as GPUs - rendering it often inappropriate for the kind of mass-production scenarios considered in this work where where products must be built with low-cost/power and high-frequency control electronics. Various works explore the possibility of augmenting traditional PID loops with a DRL-based pipeline, where the DRL agent will modify the PID parameters dynamically throughout the lifetime of a control process. These are typically referred to as adaptive or dynamic PID controllers.

An adaptive PID controller that is augmented by RL using an actor-critic method and temporal difference learning is shown in \cite{Guan2020}. Greater control stability is achieved compared to a conventional gradient-based PID tuning method is shown in a purely simulated non-linear environment. Linearity and inability to adapt to changing control scenarios are identified as major problems concerning PID controllers in \cite{Hynes2020} (in the application context of suspension control in cars). The work introduces a dynamic augmented PID controller which uses an RL agent to choose the suspension systems damping rate, using a conventionally pre-tuned PID controller at the start. It is shown to achieve marginal improvements over conventionally tuned PID controllers. Similarly, \cite{Carlucho2017} identifies key issues with PID control as being that classical tuning techniques are not entirely suitable when the system operational conditions are uncertain. A Q-learning based method is presented that implements an adaptive PID controller, incorporating temporal memory into the learning process to improve performance in a real (non-simulated) system and showing robust performance for controlling a mobile robot.

While the aforementioned methods all demonstrate a clear capability of DRL to improve PID tuning outcomes, they are generally designed on the basis of a dynamic PID control process, where the online PID control-loop is augmented in some way to incorporate the DRL elements. As noted before, with respect to minimising compute hardware complexity/energy requirements/cost and maximising control loop frequency, such methods fall short of simpler control methods like classical PID-loops. While artificial intelligence accelerator devices are becoming faster, more cost effective and energy efficient, the cheapest/most power efficient hardware used commonly in mass-manufactured products is still not ideally suited for these types of calculations.

Another adaptive PID controller is detailed in~\cite{Lee2021}, with the goal of reducing the number of tuning attempts required to find good parameters. This is comprised of a neural architecture which takes information about previous PID coefficients and performance as input and produces a new set of PID coefficients and is able to (mostly) complete tuning with 3 attempts. This is less restricted with respect to compute requirements, as the DRL-based PID tuning is done once in closed-loop device operation to generate parameters that can be used in a regular PID loop. However, the method still requires the presence and operation of the device to be optimised (i.e. it is an online tuning process, not offline) and doesn't account for how physical attributes of the underlying control system may contain useful information for determining suitable PID parameters, so must still be repeated per device.

The randomness associated with the dynamics of a control system (as is true for any real-world system) is identified in \cite{Shipman2019} as a difficulty in the face of training DRL agents from scratch that will directly change the controller output over the lifetime of a control process. A staged training process (based on a simulation of the control system) is implemented that gradually introduces the system-randomness as the agent learns a control policy over time and is compared to a DRL training procedure that is exposed to full randomness from the start. The staged training procedure DRL agent to learn slightly better policies faster. The method can also learn policies that perform similarly ($\approx 3\% worse$) than direct synthesis methods without requiring a model of the underlying control system. This work demonstrates the advantage of using DRL for PID tuning when perfect system models are unknown.

\section{Optimal Control and Optical Switching with Piezoelectronic-Actuators}
\label{section:problem}
\subsection{Optical switching using piezoelectronic-actuators}
\label{section:pea_switching}

\begin{figure*}
    \centering
    \includegraphics[width=0.4\textwidth]{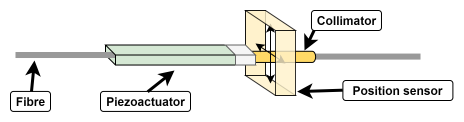}
    \includegraphics[width=0.4\textwidth]{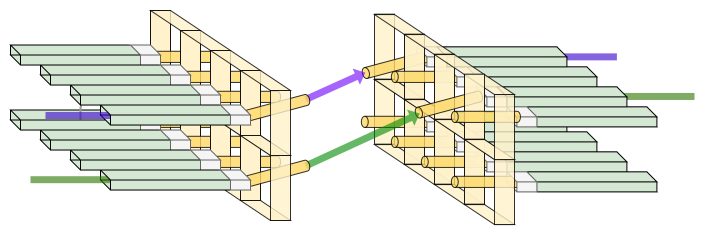}
    \caption{(Left) Diagram of a single actuator-based sub-system defining a single port in a switch. Data from the position sensor can be used in a closed loop alongside x- and y- axis driving voltage to implement PID control on the position of the port. (Right) Diagram of an input and output plane based on actuators as used in actuator-based optical switch design. Input ports and output ports can point towards each other to create viable light paths and do not require light to be passing through the system to do so.}
    \label{fig:port}
\end{figure*}

Optically switched data centre (DC) networks can provide significant benefits over electronically switched ones, particularly in the current age of extreme growth in Cloud based services. Optically switched DC networks can provide security (the network is passive so can't interact with transmitted data), reliable performance guarantees, increased server-server bandwidth, as well as opening routes to entirely new disdaggregated DC architectures, which can give way to both more efficient resource utilisation as well as more flexible resource pooling options~\cite{shabka2021resource, Shabka20, Mishra21}. Optical circuit switched network backbones are also shown in hyperscale sized data centres to reduce power consumption and increase capacity \cite{Jupiter}. A common type of optical switch is based on free-space beam-steering between input and output planes, where the steering is implemented with a piezoelectric-actuator undergoing a control process \cite{Deakin19}. Piezoelectric-actuators are small electro-mechanical devices whose planar movement is controlled by applying a particular voltage along each of its two degrees of freedom. 

Each actuator is controlling the direction that a collimator lens (that has a fibre directing a beam of light into it) is pointing, as shown in the left hand side of Figure~\ref{fig:port}. Switches can be built by constructing an input plane and output plane, each with $M\times N$ actuators. By moving the position of a particular input-output pair of actuators such that they are pointing towards each other (referred to from now on as a `switching event'), a light path can be established. The right hand side of Figure \ref{fig:port} visualises how these input/output planes can be constructed as well as the notion of light paths between the input and output plane being established on the basis of actuator position control.

Ports are constructed with a position sensor per degree of freedom for each actuator, such that the position of the actuator can be measured in real time. As such, closed loop control processes based on targeted positions (such as PID control) are implemented in order to ensure the stability, reliability and speed of these switching events. It is also noted here that an active optical signal is not required to establish these light paths, since the positions are controlled by voltages supplied independently of the optical signal. This is what is referred to as `dark switching'.

\subsection{Piezo-electronic actuator control requirements}
\label{section:control_requirements}

The performance features mentioned in section~\ref{section:pea_switching} are described more formally below. The following 3 performance features described will serve as the 3 performance metrics evaluated throughout this work.

\paragraph{\textbf{Settling-time ($ST$)}} This is defined as the amount of time taken for the position of an actuator to fall and remain within a particular margin, $\pm ST_{margin}$, of the desired position where this margin is set relative to what is practically relevant (e.g. a very small margin corresponds to a very low port-to-port optical loss ). $ST$ must not be larger than a particular target value, $ST_{max}$. Minimising $ST$ equates to minimising switching overhead and therefore minimising communication latency incurred due to switching. For the work presented here, $ST_{max} = 20ms$ and $ST_{margin} = \pm 0.15mm$. We also note that in the commercial domain, for the very large switches used in this work, $ST_{max} = 50ms$, and the smaller 20ms value applies to the smaller switches. Since travelling smaller distances requires less extreme corrections under control, smaller switches (e.g. $16\times16$ can generally switch faster than larger ones. In this work we observe the more stringent 20ms settling objective while working with a larger switch. This allows us to address a more difficult switching scenario and show that even at the limit of more stringent product performance specifications, deep reinforcement learning can be employed to design even better performing switches.

\paragraph{\textbf{Overshoot ($OS$)}} This is the maximum distance away from the target that the actuator is at any point during switching. Its absolute value must be smaller than a particular limit, $OS_{max}$. If overshoot is sufficiently large, then the switching port's input/output position may `leak' into an incorrect destination, compromising the accuracy of the switch and introducing cross-talk induced noise as well temporal increased insertion loss and loss fluctuation. For the work presented here, $OS_{max} = 5mm$

\section{One-step PID Tuning with Deep Reinforcement Learning}
\label{section:solution}
This work presents a method for finding optimal PID parameters based on DRL that overcomes the difficulties noted in section \ref{section:intro}. The method achieves both a control performance improvement, whilst also significantly reducing optimisation overhead. This section will detail the relevant background to and formulation of this method, and highlight it's benefits at a high level. Following this,  experimental results are detailed in section \ref{section:results}. 

\subsection{Deep Reinforcement Learning}

Reinforcement learning refers to any problem consisting of an \textit{environment} and an \textit{agent} which interacts with that environment by taking actions within it. Generally, a solution to a reinforcement learning problem is to find a behavioural policy that the agent can follow such that some outcome is maximised over the long term (e.g. find a control policy so that an agent controls a helicopter to remain stable for as long as possible). This is typically referred to as the \textit{optimal policy}. An agent will observe the state of the environment, determine what action to take based on this information and will then receive some reward on the basis of the outcome of that decision. A reinforcement learning algorithm will iterate through this cycle many times, making updates to a policy based on the received reward in order to move it closer to the optimal one.

More formally, the environment is represented by a tuple $<S,A,R_a,T_a\gamma>$ called a \textit{Markov Decision Process}. The items in this tuple are:
\begin{itemize}
    \item $S$ [state space]: the set of all possible states that the environment can be in.
    \item $A$ [action space]: the set of all possible actions that an agent can take in the environment.
    \item $R_a(s,s^{'})$ [reward function]: the reward obtained from being in state $s \in S$, taking action $a \in A$ and ending up in state $s^{'} \in S$.
    \item $T_a(s,s^{'})$ [transition function]: the probability of being in state $s \in S$, taking action $a \in A$ and ending up in state $s^{'} \in S$.
\end{itemize}
MDPs are also often episodic, meaning that there is some terminal state after which no further different states can be visited. For example, a checkmate position in a chess game MDP is a terminal state, since no further states can follow that one. The environment must then be reset (e.g. put the chess pieces back to their starting positions and reset the clocks) before further interaction can continue.

Finally, a policy is defined as a some function $\pi: S \rightarrow A$ which maps states to actions. The standard policy update loop, defined in the context of the problem addressed in this work, is visualised in Fig. \ref{fig:experiment}.

Broadly speaking, reinforcement learning algorithms fall into two groups; policy gradient and q-learning. Policy gradient methods approximate the optimal policy function, $\pi(s)$ directly, whereas q-learning methods approximate a function, $Q_{\pi}(s,a)$, estimating the reward achieved by the optimal policy given that you are in state $s \in S$ and take action $a \in A$ before following policy $\pi$ until end-of-episode. Actor-critic methods combine both of these methods, using the approximation of $Q_{\pi}(s,a)$ to update $\pi(s)$ based on an estimate of the outcome of it's action. When $\pi(s)$ and/or $Q_{\pi}(s,a)$ and approximated with a deep neural network - as in this work - this is typically referred to as \textit{deep reinforcement learning}. Detailing of the more fundamental technical description of reinforcement learning will not be provided here, but is clearly described in \cite{rl_lectures}.

As will be seen, an action in the context of this work consists of choosing several continuously valued numbers from a continuous set. As such, q-learning methods - which can only support discrete action spaces - are not relevant if the continuity of the action space is to be maintained (as opposed to discretising it and therefore restricting possible actions). The Proximal Policy Optimsiation actor-critic algorithm (PPO) was chosen due to it's sample efficiency, ease of hyperparameter tuning and generally good performance across a range of tasks in the continuous action space domain \cite{SchulmanWDRK17}.

\subsection{One-step offline PID optimisation as a MDP}
\label{section:mdp}
Here we specify the MDP used to formulate the PID optimisation as a DRL problem. The environment in this scenario is a PID-controlled system, which in the case of the presented experiments is an actuator in an optical switch. A DRL agent learns to make one step optimisation decisions to determine the best set of PID parameters (for both axes simultaneously) based on basic physical state information about the actuator and guided by a multi-objective training regime. This process as described below is visualised in Fig. \ref{fig:experiment}

\paragraph{\textbf{State/Observation}}: The state of an actuator is represented as information about the resonance frequency, $\omega$, and gain, $G$ properties (per axis):

\begin{equation}
    s = [\omega_{x},G_{x},\omega_{y},G_{y}]
\end{equation}

Feature scaling is applied to the state. The raw measurement value of each feature type (i.e. frequency and gain) have different order of magnitudes to each other, though all measurements of the same feature type share the order of magnitude (i.e. all resonance measurements are the order of $10^a$ and gain of the order $10^b$). As such, each feature is normalised with respect to it's type's order of magnitude. This ensures that all features vary at the order of $10^1$ and that no feature is disproportionately influential during training.

The actuator state is measured each episode rather than using the pre-determined measurement done during factory characterisation. This is done to avoid overfitting to a set of pre-measured state values, especially relevant if a small number of actuators are used for training. The error inherent in measurement processes will expose the agent to a more realistic distribution of states an should allow it to learn a policy that accounts for this uncertainty. As such, it generates parameters in production based on some pre-measured state of a device, it should output parameters in a way that accounts for the uncertainty in that state measurement process too. By contrast, on testing/inference the agent generates parameters for previously unseen devices based on factory-characterisation measurements that induce very good control performance, showing that the policy is indeed generalised across devices.

Additionally, both axes are shown at once to the agent. Since each axis is controlled by distinct (non-coupled) control loops, it could be conceivable that an agent learns to output PID parameters for each axis separately. However, as noted in section \ref{section:desirable_features} some inter-axis coupling effects present in the system may be subtle and difficult to measure amidst other more prominent sources of instability. By allowing the agent to control both axes simultaneously, should coordinated axis-parameter values lead to a better reward during training, then this aspect of the policy can be reinforced. As such, by allowing the agent to interact with both axes simultaneously, sufficient information should be available to the agent via the reward signal for it to mitigate these effects with suitable policies. This premise is continued below in the discussion of the benefits due to the action space design.

\paragraph{\textbf{Action}}: We define the action space as the space of all possible PID coefficients (for both axes simultaneously). 
\begin{equation}
    a = [P_{x},I_{x},D_{x},P_{y},I_{y},D_{y}]\ where\ \{P, I, D\} \in \mathbb{R}
\end{equation}
where $P_{x/y},I_{x/y},D_{x/y}$ are the proportional, integral and derivative parameter values for the $x/y$ axis respectively. In practise, an approximate range or order of magnitude for the P, I and D parameters will be known for the device; a control system will typically have some associated ranges for each PID parameter, outside of which control behaviour may be divergent. If not known ahead of time, since an infinitely large action space consisting of all real numbers is intractable, a rough iteration of a simple procedure such as that presented in \cite{ziegler1942optimum} can be implemented in order to determine the order of magnitude of each parameter. Beyond this, no further restrictions are required to be imposed on the action space.

Similarly to the above, the PID values for both axes are output simultaneously. While inter-axis effects may be difficult to measure directly, they could still be present and disruptive to the stability of the actuators position. If there was some disruption due to this phenomenon, this would be reflected in the reward signal received by the agent in the simple form of a larger penalty. As such, being able to see state information about each axis, as well as simultaneously make changes to each of their control loops, the agent can in principle respond to these effects simply by optimising the policy with respect to the reward signal as normal. If state information and PID parameter outputs were separated by axis, then the agent would not be able to relate reward penalties to inter-axis coordination in it's policy and performance would suffer. On the other hand, with the current state- and action- space form, if no such coupling effects are present, the reward function will not have any additional penalty, and so the policy is free to effectively modify each control loop separately.

\paragraph{\textbf{Reward}}: In this work we use a scalar multi-objective reward function to influence an agent to address the multiple relevant metrics for a given control process. Given a set of metrics (e.g. settling-time, overshoot etc), $S$, we define the measured value of a metric $s \in S$ after a actuator has enacted a switching process with some set of PID parameters as $m_s$. Moreover, for each metric $s \in S$ the user will determine an acceptable threshold for that metric, $l_s$, such that the DRL agent will be punished should their PID parameters drive a switching event that yields a measurement $m_s > l_s$. These thresholds are generally informed by the context of the application. We implement this in the following way:
\begin{equation}
    R = -\prod_{i} f(m_i,l_i)
\end{equation}
where
\begin{equation}
    f(m_i,l_i) = 
    \begin{cases}
        \frac{m_i}{l_i},& \text{if } m_i\geq l_i\\
        1,& \text{otherwise}
    \end{cases}
\end{equation}

In taking the product of each metric's contribution to the reward, a more punitive multi-objective criteria is imposed than other aggregations such as addition. This is becuase it is not possible to `cheat' by finding a policy that optimises some metrics at the expense of others, since the product will still be large if some metrics are significantly over their threshold.

In our experiments we use the metrics $S = \{ST,OS\}$. Additionally, since there is a reward value associated with each axis, we use the axis with the worst result:

\begin{equation}
    R = -\prod_{i} max(f(m_{ix},l_{ix}),f(m_{iy},l_{iy}))
\end{equation}

In this way, both with respect to the observed metrics and actuator, the reward function incentivises a policy which minimises the worst case performance with respect to axis and metric. This is designed in part to reflect the generic requirement of mass-manufactured products which in general must provide some sort of guaranteed service/performance level to users.

This also allows for a simple means of asserting priority to certain metrics. Rather than setting the threshold to precisely the acceptable limit, it can simply be set lower such that the metric generates a reward penalty even when it is within acceptable working range but still above a more aggressive lower threshold. This is useful in scenarios where one out of two contending metrics is more important to maintain close to or below threshold. For example, if metrics $A$ and $B$ have thresholds $X$ and $Y$ respectively, where $X < Y$, then when both metrics exceed their respective thresholds by the same percentage, $A$ will contribute more to the reward signal than $B$.

\paragraph{\textbf{Transition/Episode}}: Episodes are one-step, where the agent receives an initial state, 
\begin{equation}
s = [\omega_{x},G_{x},\omega_{y},G_{y}]
\end{equation}
and takes a single action
\begin{equation}
a = [P_{x},I_{x},D_{x},P_{y},I_{y},D_{y}]
\end{equation}
before receiving a reward based on the response of that actuator when it undergoes a random switching event using those parameters. At the end of the episode, a new actuator and switching event is randomly selected from the set of training actuators.

Choosing a new random actuator and switching event each episode allows for a broad statistical exposure of the agent to the full range of switching events and actuator variation over the lifetime of it's training. Over enough training episodes, the agent should reasonably have been exposed to an approximately accurate distribution of switching events with respect to the full range that might be experienced throughout the average switch's lifetime. Since it will have also been exposed to the underlying distribution of actuator states (from repeatedly measuring those in it's training set) and will have received numerous rewards for various PID parameters used to switch these actuators through a variety of switching events, it should in principle be able to learn a generic policy that can map an actuator state to a set of PID parameters that will yield good switching performance on average over the full range of possible switching events.

\section{Experiment}

In this section an experimental setup is described. This setup was used to train and test the proposed DRL-based parameter optimisation method and compare it against the default baseline method.

\subsection{Test-bed Setup}

\begin{figure*}
    \centering
    \includegraphics[width=0.7\textwidth]{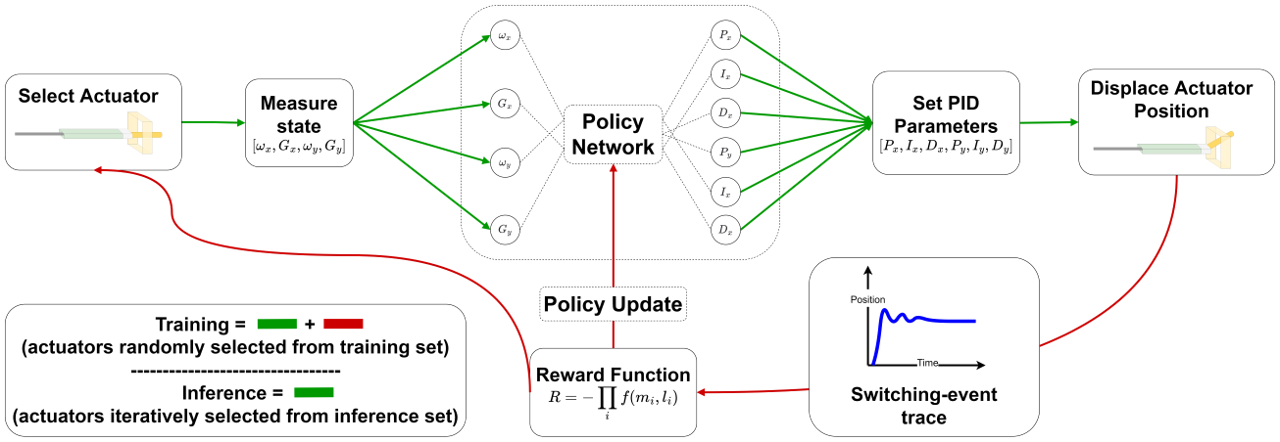}
    \caption{Flow diagram showing the training and inference loops implemented. The training follows the full cycle of the green path followed by the red path, whereas the inference process follows only the non-cyclic green path to produce PID parameters for a single actuator.}
    \label{fig:experiment}
\end{figure*}

Our experimental test-bed includes a PC, a set of actuators, and a controller used to interface with the actuators (e.g. send switching commands, trace position etc). This setup and the associated training and testing loops are visualised in Figure \ref{fig:experiment}. The set of actuators + controller are effectively a deconstructed optical switch since the components are identical to those that are found in a real optical switch product. The PC is used to implement the DRL policy and all training and testing algorithms, as well as post-processing of actuator signal traces to calculate the rewards/performance metrics. The controller is effectively used as a pass-through device to interface between the policy output and the PID values of the actuator, as well as to send instructions to and receive data from the actuators.

Each port is operated in closed-loop PID control, where the control loop is implemented on an FPGA at 10kHz simultaneously for all ports in a switch. In our experiments, we use a set of 32 actuators during training of the RL model, where each actuator can switch to 384 unique positions as it could if it were part of a $384\times 384$ port optical switching product. This is done since this is the largest switching product available, and as such ensures that the actuators will on average experience the most extreme operational conditions due to bigger movements required to access the further away ports. This means that the training/testing scenarios are as tough as possible, since smaller switches undergo a less extreme set of swtiching events and therefore have smaller $ST_{max}$ as referenced in section \ref{section:problem}. An additional set of 16 actuators are reserved for testing and the agent does not interact with these during training.

The high level structure of a actuator-based port, as well as a visualisation of how they are built into an optical switch is shown in the left and right hand image on Figure \ref{fig:port} respectively. The direction the actuator will point to in 2D space (referred to now as it's position) is controlled by voltages applied to the actuator along the X and Y axes separately. An optical fibre is attached to each actuator. At the end of the actuator there is a rod with a collimator lens at the tip, which is used to focus the optical signal propagating into/out of that port's fibre. In this way, the actuator is used to direct the propagation of any light through that ports fibre. Surrounding this rod is a position sensor from which a digital reading of the ports position along each axis can be read separately. Therefore, a closed-loop dynamic control process (e.g. PID control) can be implemented per-axis between the actuator driving voltages and the position sensor when a particular target position is known. This basic port architecture is visualised in the left hand image of Figure \ref{fig:port}.


\subsection{Baseline}
\label{section:sota}

The baseline PID tuning method used as a baseline is referred to as the \textbf{default}. This method is based on direct-search (i.e. exhaustively searching numerous parameter combinations until some objective is accomplished). Since no two actuators are identical, technically a fully exhaustive search procedure would be performed uniquely once per actuator, since parameters optimised on the basis of one actuators control performance do not necessarily guarantee the same outcomes for another. There are also many different control movements (i.e. position $(x_1,y_1)$ to $(x_2,y_2)$) that an actuator can undergo as discussed in section \ref{section:intro}, and they are not necessarily all controlled equally well by a given PID parameter combination. Finally, since PID parameters are 3 continuous values, the search space must be confined to make it tractable under direct search (as similar noted in section \ref{section:solution}).

To achieve this, several augmentations to the fully exhaustive direct search method are implemented. Firstly, a set 16 of extreme switching events are chosen to be tested. Specifically, these are 8 events spanning each of the 4 centre-to-corner, as well as all 4 opposite corner-to-corner movements. These represent the most extreme movements that a port will undergo from centre position to destination, and in absolute distance respectively. The motivation behind this design choice is that tuning with respect to these extreme switching events should optimise with respect to worst case performance, and therefore also keep the less extreme switching events below threshold for all metrics too.

A set of 32 actuators are used to tune PID parameters, where parameter quality is judged by ST (provided that OS is still less than the allowable maximum). When the average performance of the set of tuning actuators is evaluated, it can be tested on a separate set of 16 actuators. This ensures that the tuning process finds parameters that are at least somewhat general with respect to actuator variety.

PID values are iterated over grid search, where suitable starting points for each of the three parameters can be determined by simple and well known methods \cite{ziegler1942optimum}. Grid search is employed with 6 decimal place precision, centred at these starting points. When a set of parameters is found to yield acceptable (with respect to the aforementioned ST and OS targets) performance, they are applied to the additional set of actuators. If the performance remains consistent, this parameter combination is considered viable, and the best of all viable combinations will be selected after tuning as the single generic parameter set. Following this procedure, this same set of parameters is used for all actuators.

\subsection{Training}

We train a DRL policy network directly using the well-established PPO algorithm \cite{Schulman2017}, implemented using the RLlib \cite{Liang2017} and PyTorch \cite{Paszke2019} libraries for reinforcement and general deep learning respectively. After testing across various policy network sizes, a network with two 16-unit hidden layers was used. Training is stopped after the reward is convergent, or after 1600 episodes (since it was found that policies generally converged within this limit).

The training loop is visualised with the full cycle defined by the green arrows followed by the red arrows shown in Figure \ref{fig:experiment}. Similarly, Algorithm \ref{alg:algorithm} details the training process where $\pi_\theta$, $\omega$ and $G$ are the DRL policy parameterised by parameters $\theta$, actuator resonanace and actuator gain respectively, $P_i$, $I_i$ and $D_i$ are the proportional, integral and derivative parameters respectively for the $i^{th}$ degree of freedom of an actuator ($i \in \{x,y\}$), $R$ is the reward function and $l_{j}$ is the reward threshold for metric $j$. The metrics accounted for in the reward function are settling-time ($ST$) and overshoot ($OS$).

\begin{algorithm}
\caption{Pseudo-code showing the high level PPO-based training process undertaken by the DRL agent.}\label{alg:algorithm}
\begin{algorithmic}
\State choose values for $l_{ST},l_{OS}$
\For{$\_ \in number\ of\ batches$}
    \State $batch = [\ ]$
    \For{$\_ \in batch\ size$}
        \State $actuator \gets random\ actuator\ from\ training\ set$
        \State $dst \gets rand om\ switching\ destination$
        \State $s = [\omega_{x},G_{x},\omega_{y},G_{y}]\gets actuator\ state\ measurement$
        \State $a = [P_{x},I_{x},D_{x},P_{y},I_{y},D_{y}] \gets \pi_{\theta}(s)$
        \State set PID parameters for $actuator$ with $a$
        \State $m_{ST},m_{OS} \gets$ $actuator$ switch to $dst$
        \State $R \gets R(m_{ST},m_{OS},l_{ST},l_{OS})$
        \State add $(s,a,R)$ to $batch$
    \EndFor
    \State update policy parameters $\theta$ with PPO update rule
\EndFor
\end{algorithmic}
\end{algorithm}

\subsection{Testing}

We compare the performance of the learnt policy to the default method across all metrics and in several different scenarios.

The standard test run consists of observing the switching performance of 100 pseudo-random switching events, where source actuator and destination position are randomly chosen each time, as in the training scenario. This same set of random switching events is used for all tests on both methods for consistency. Where in normalised time the actuators are able to change position at the order of $\mathcal{O}(10ms)$, a switching trace of $\mathcal{O}(100ms)$ is recorded in order to observe the long-term stability dynamics of the actuator as well as the initial movements associated with switching. This is to ensure that a policy does not switch actuators very quickly, but also lead to divergent behaviour in the long term.

The switching process is controlled by PID parameters which are determined by the method being tested. The testing loop is visualised in Figure \ref{fig:experiment} by following the green arrows only. The process is repeated over 100 testing episodes using the set of test actuators, but is represented as non-cyclic since there is no feedback loop involved as is the case for the training process (which follows the red arrows after the green).  Across all tests detailed in this section, the state of the actuators in the test set were measured once at room temperature and then not again throughout all testing. This is to demonstrate the viability of the method specifically in it's offline inference scenario, as would be the case were it to be applied in a real production environment as described in sections \ref{section:problem} and \ref{section:solution}. We implement the two tests stated below.

\textbf{General performance test}: In this test the switching performance with respect to settling time and overshoot are shown as examined at room temperature. The methods are co-examined along both absolute performance, as well as performance consistency and effectiveness with respect to product requirements. We also examine how each method performs when more punitive switching performance requirements are imposed. Comments are also made regarding in-production efficiency benefits.

\textbf{Temperature sensitivity test}: We also examine the resilience/sensitivity of the switching performance across a range of temperatures. For the RL method, actuator states were measured once at room temperature ($25\degree$) and used for all other temperatures. We repeat the experiments shown previously for each of the following temperatures: $5\degree C,\ 25\degree C,\ 35\degree C,\ 53\degree C,\ 73\degree C$. These experiments are done to observe the variability of the statistical performance of each method under thermal variation. Ideally the performance of some method will be stable across the full range of temperatures, since optical switches are used in both cold environments (e.g. chilled server-rooms, outer space) as well as hotter environments (e.g. portable television infrastructure).

\section{Results}

The experiments are carried out on real optical-switching products manufactured by a world-leader in this domain. For this reason, the results presented in this section can be interpreted as more than a proof-of-concept and in fact demonstrate that the proposed method is truly viable in real-world commercial production environments where large numbers of controlled devices/components are manufactured and used in products which have strict performance requirements.

\label{section:results}

\subsection{General performance evaluation}

\begin{figure}
    \centering
        \includegraphics[width=0.55\textwidth]{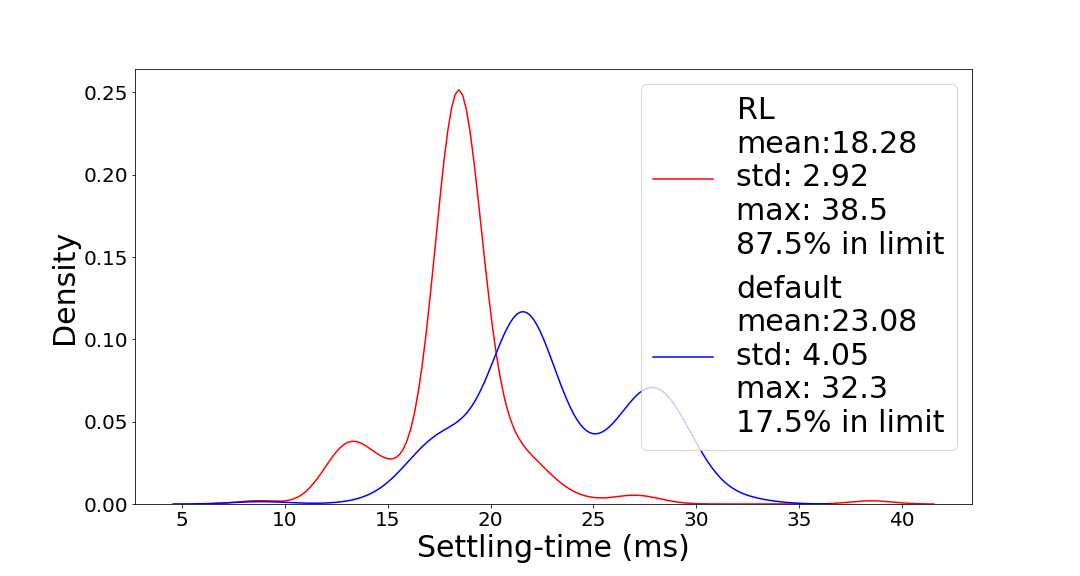}
        \includegraphics[width=0.4\textwidth]{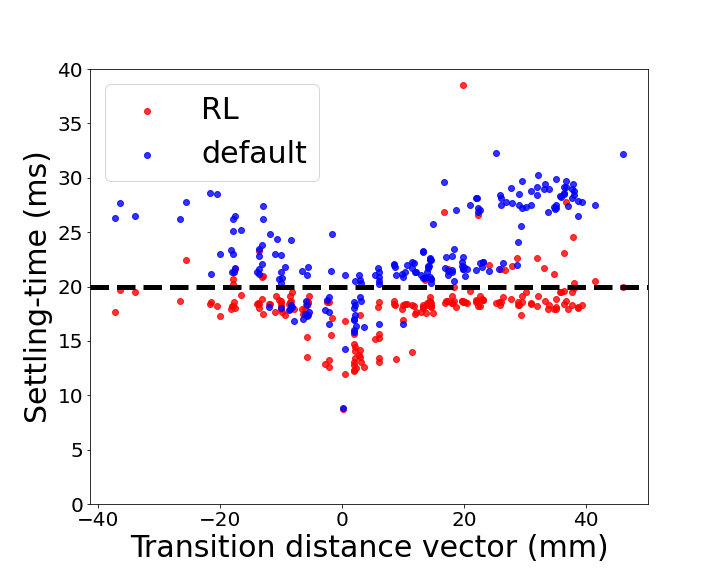}
    \caption{(Left) Distribution of settling time measurements for each of the test switching events (per-axis) for DRL and defualt methods. (Right) Scatter plot showing the measured settling time for each switching event (per-axis) plotted against the corresponding transition distance required by that switching event. Black dotted line denotes the target settling time, below which each settling event would ideally be.}
    \label{fig:settling}
\end{figure}

\begin{figure}
\centering
        \includegraphics[width=0.55\textwidth]{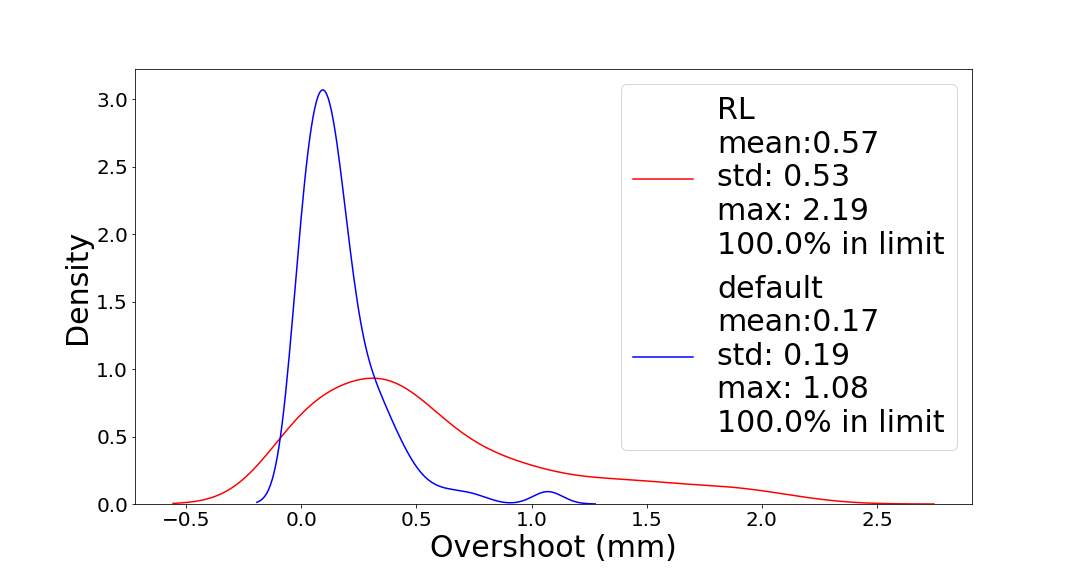}
        \includegraphics[width=0.4\textwidth]{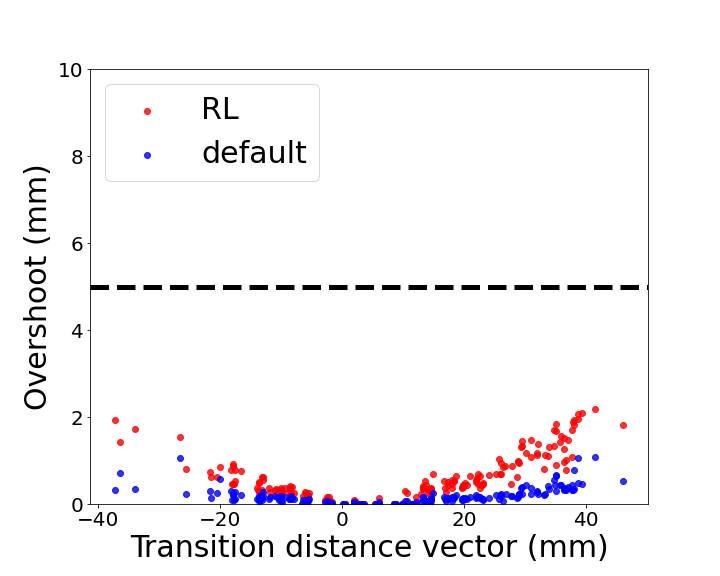}
    \caption{(Left) Distribution of overshoot measurements for each of the test switching events (per-axis) for DRL and defualt methods. (Right) Scatter plot showing the measured settling time for each switching event (per-axis) plotted against the corresponding transition distance required by that switching event. Black dotted line denotes the target overshoot, below which each settling event would ideally be.}
    \label{fig:overshoot}
\end{figure}

Re-iterating performance objectives, it is required that the ST values are minimised and ideally equal to or below the ideal target time, while OS values are required to not exceed a particular value. As such, in this regular operational scenario $ST$ is the `prioritised' metric since performance does not increase further once $OS$ is within it's threshold, whereas it is always desirable to reduce $ST$ as much as possible. As such, we the agent's tested here have been trained with the ST reward threshold set to 0 (i.e. always punish proportionally to absolute measured ST values) whereas the OS threshold is simple the production requirement (5mm). 

Figures \ref{fig:settling} and \ref{fig:overshoot} show the distribution of results for ST and OS respectively achieved by each method. These plots show that the DRL method ensures that $5\times$ more switching events settle within $\pm0.15mm$ of the target position at or below the ideal target time of 20ms (87.5\% vs 17.5\% for DRL and default respectively). Additionally, it also achieves a 22.2\% reduction in the mean ST value, reduces the worst case ST value by 28\% and has a 43\% smaller standard deviation. While $OS$ performance is indeed worsened by the DRL method, this is acceptably so since the overshoot values are still 100\% under the acceptable threshold for this metric so do not violate any performance requirements. These results show very clearly the significant benefits that the DRL method can provide, in a testing scenario that accurately represents how this methodology would be applied in production. It is seen that the DRL method is able to acceptably sacrifice one metric ($OS$) for major improvements in a higher priority one ($ST$), without incurring an unacceptable penalty with respect to the sacrificed $OS$ metric.

\begin{figure}
    \centering
        \includegraphics[width=0.45\textwidth]{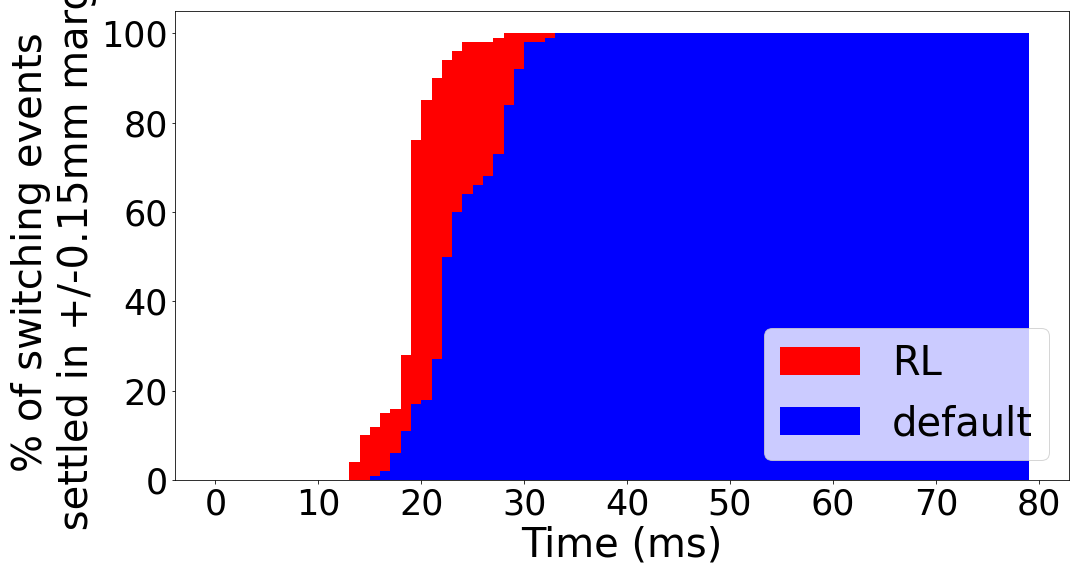}
        \includegraphics[width=0.45\textwidth]{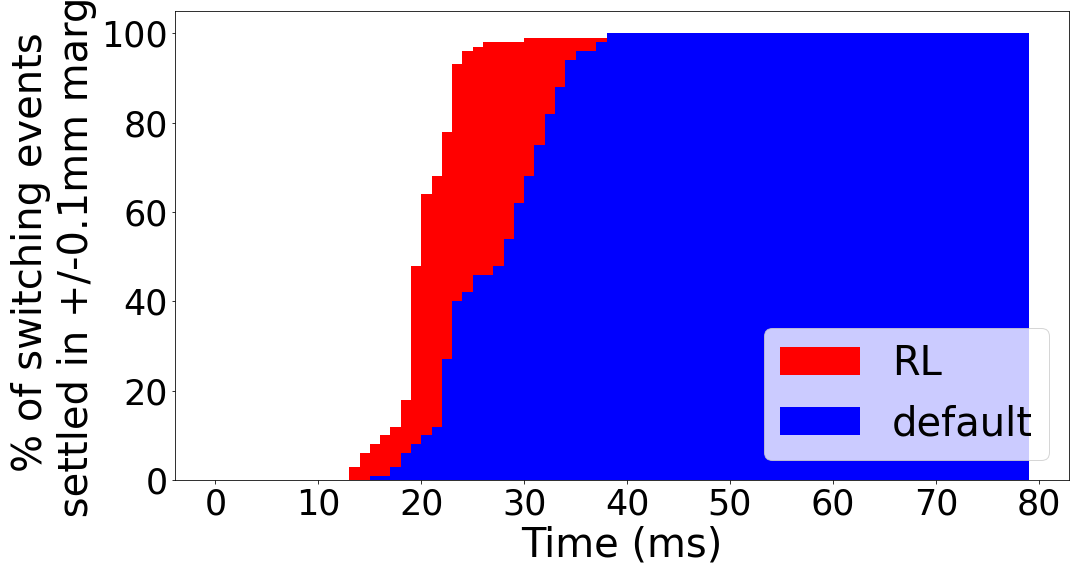}
        \includegraphics[width=0.45\textwidth]{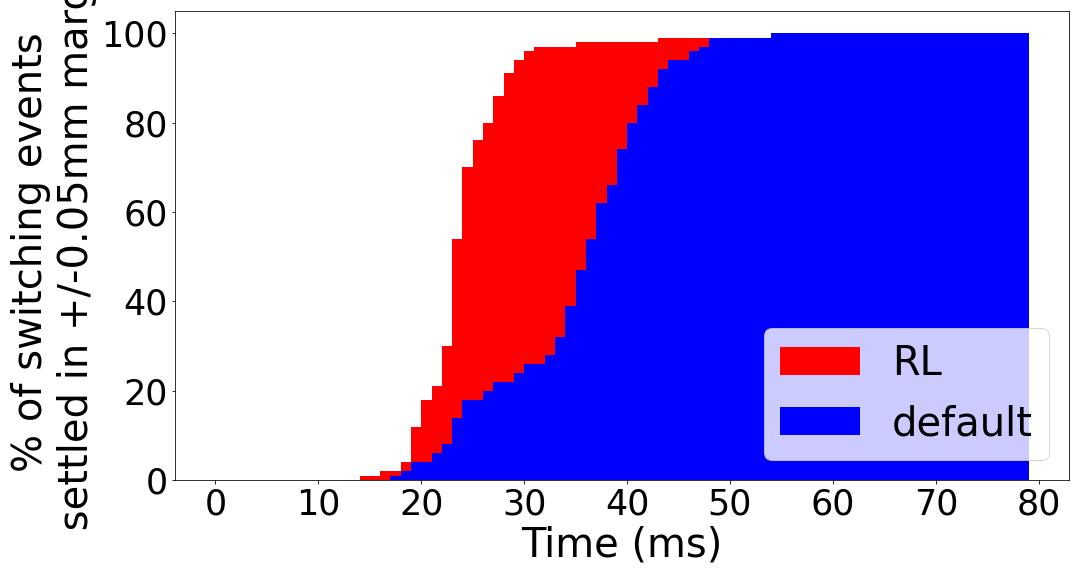}
        \includegraphics[width=0.45\textwidth]{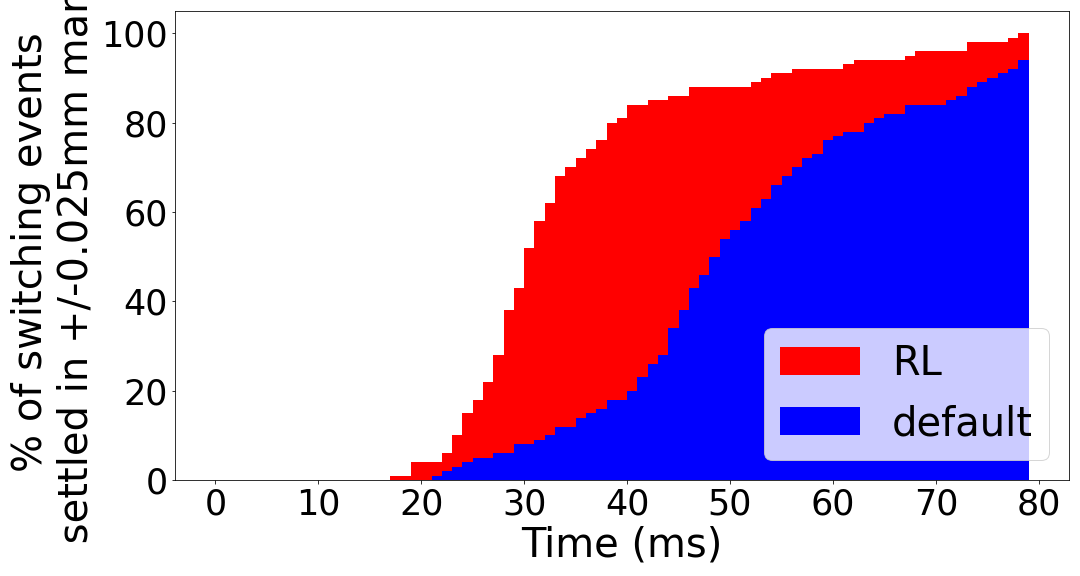}
    \caption{CDF of \% of test switching events that are in-margin for margins of (clockwise from top-left) $\pm0.15mm$, $\pm0.1mm$, $\pm0.025mm$ and $\pm0.05mm$.}
    \label{fig:cdf1}
\end{figure}

\begin{figure}
    \centering
        \includegraphics[width=0.45\textwidth]{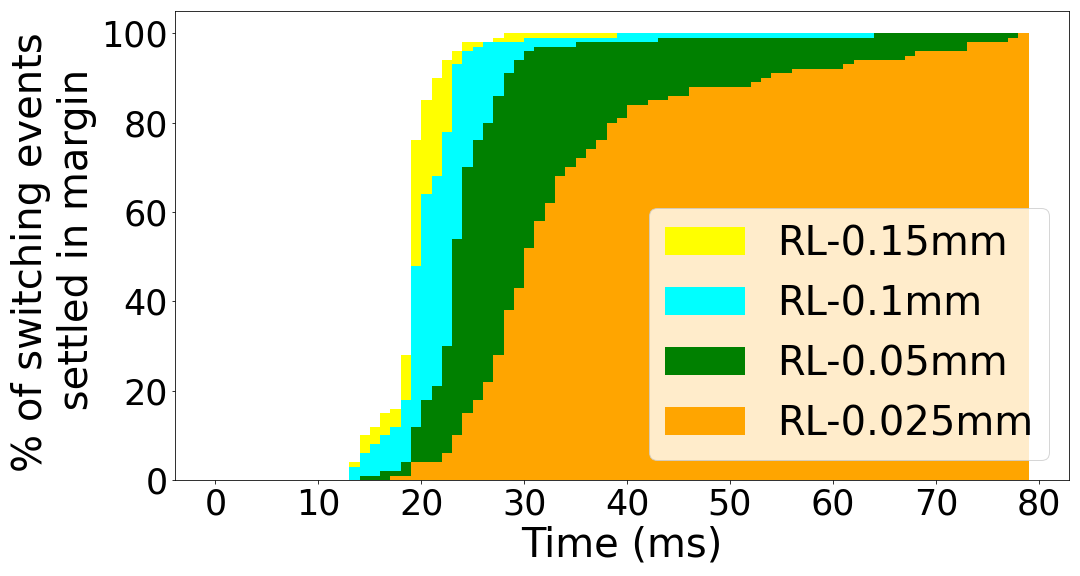}
        \includegraphics[width=0.45\textwidth]{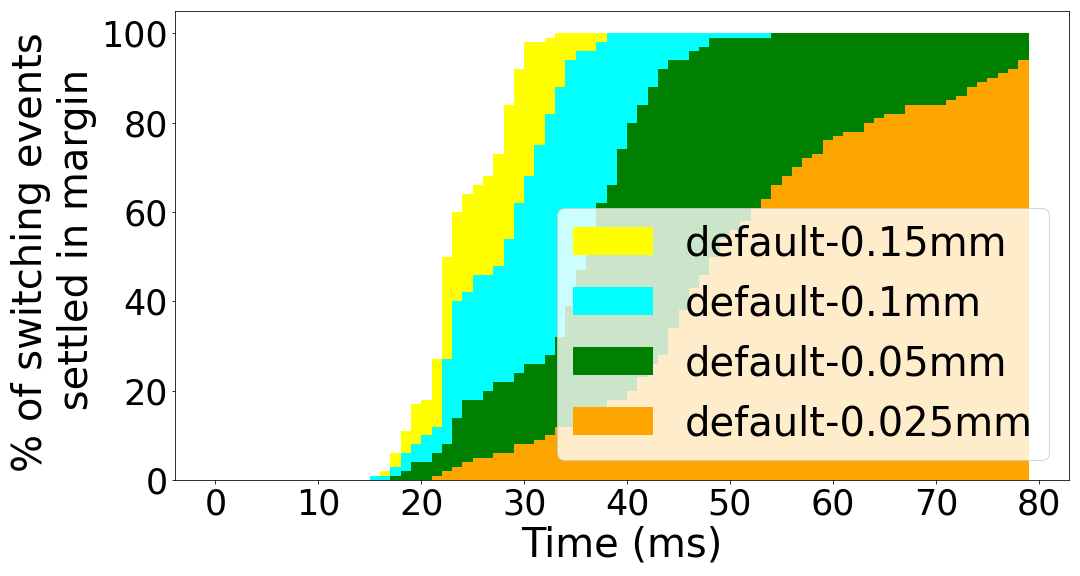}
    \caption{CDF of \% of test switching events that are in-margin for margins of $\pm0.15mm$, $\pm0.1mm$, $\pm0.05mm$ and $\pm0.025mm$ for the DRL method (left) and the default method (right).}
    \label{fig:cdf2}
\end{figure}

The 4 images of Figure \ref{fig:cdf1} show a CDF plot visualising (over a set of test switching events) how many of those events had settled within a particular margin after a certain amount of time, comparing those results for a trained DRL agent and the default parameter set. Going clockwise, the respective margins in each image are $\pm0.15mm$, $\pm0.1mm$, $\pm0.05mm$ and $\pm0.025mm$, where as before 0.15mm corresponds to the target settling margin and the others are smaller more punitive margins. The upper right image (corresponding to the target settling value) expands on the results stated earlier and shown in Figure \ref{fig:settling}, indicating that along side 87.5\% of events settling in-margin by the target time, this becomes 95\% after 3 additional ms pass (15\% overhead relative to the target time). By contrast, the default method takes an additional 10ms (50\% relative to the target value) to get to 95\% settled. The other 3 images show the same analysis applied to smaller settling margins. These additional plots show that as the margin decreases, the gap between the default and DRL methods increases considerably. This indicates that the DRL-optimised parameters stabilise more quickly in general, rather than just with respect to a particular margin. 

Building on this comparison, Figure \ref{fig:cdf2} compares the CDF of each method with itself at the 4 observed settling margins. For the 0.15mm, 0.1mm, 0.05mm and 0.025mm margins respectively, the DRL agent settles 95\% of events in 23ms, 24ms, 30ms and 67ms. By comparison, the same measurements for the default parameters yield 30ms, 35ms, 41ms and $>$80ms respectively. This shows clearly that the deterioration of settling rate with respect to settling margin size is greater for the generic default parameter set, which is worse at handling more stringent margins that the DRL parameters. As such it is evident that the DRL method is yielding an all around more stable switching performance.

In addition to raw performance improvement, the DRL-based method provides major benefits in the context of production/manufacturing practicalities. Since the calculation of PID parameters for a single actuator using a trained DRL agent takes $\leq\mathcal{O}(10ms)$, the production overhead incurred by applying the DRL method to generate unique and more optimal PID parameters for each actuator in a 384-port switch is $\leq 1s$. For comparison, the default optimisation process can take the order of a week and produces only a single set of PID parameters which have been seen in this section to be more inconsistent with respect to switching performance as a result. The training procedure (which only needs to be implemented once) is $\mathcal{O}(10^2)$ faster than the direct search based tuning method. Even more compelling is the speed up for calculating new parameters when the DRL method has already been trained. This provides an absolute speed up of $\mathcal{O}(10^6)$ ($\mathcal{O}(days)$ vs $\leq\mathcal{O}(10ms)$ and can also be implemented offline and in parallel to some other process (e.g. transporting the actuators from factory line to production plant). As such the proposed method provides compelling advantages when used in production capacity.

\subsection{Temperature sensitivity test}

\begin{figure}
    \centering
        \includegraphics[width=0.29\textwidth]{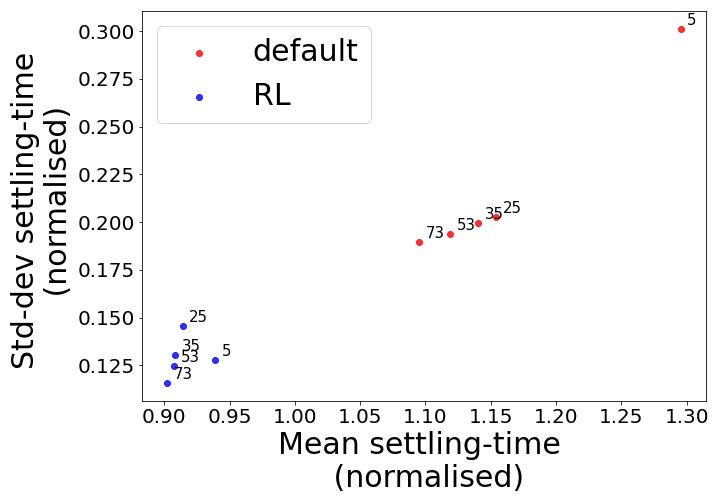}
        \includegraphics[width=0.29\textwidth]{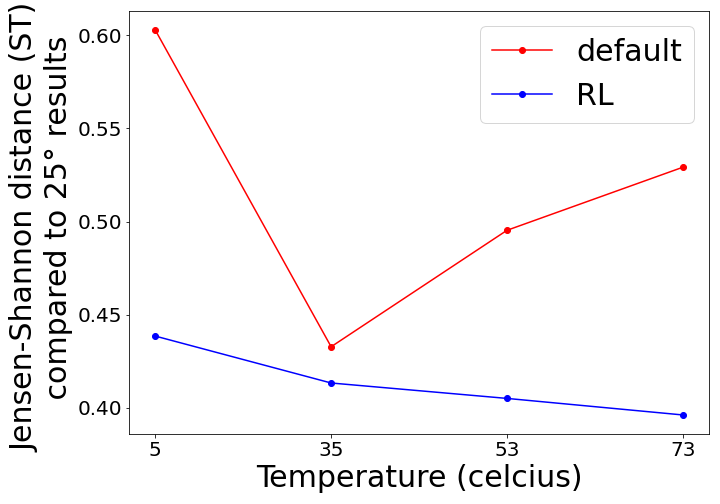}
        \includegraphics[width=0.29\textwidth]{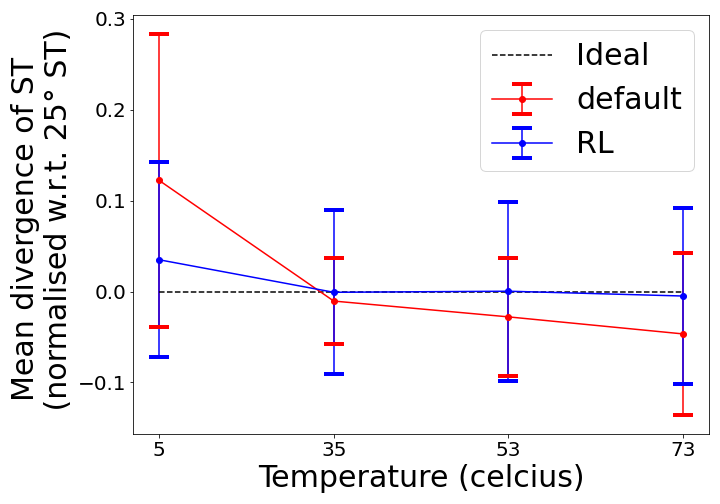}
    \caption{(Left) Scatter plot showing the mean and standard deviation settling-time for the same set of test events measured at each temperature. (Middle) Jensen-Shannon distance between the distributions of settling-time results as tested at 25\degree (room temperature) and each other tested temperature. Lower value means more similar distributions (i.e. smaller `distance' between them). (Right) Mean and standard deviation of the per-event settling-time difference between the 25\degree test and all other temperatures. Since we seek to minimise thermal variation, minimal difference (i.e. mean and standard deviation both equal 0) is desirable.}
    \label{fig:jensen_shannon}
\end{figure}

We also seek to demonstrate that the DRL-based method has more consistent performance under significant changes in operating temperature. Figure \ref{fig:jensen_shannon} presents this stability in three ways. The left-most image shows simply the mean and standard deviation of the ST of all switching events per temperature, where each temperature is tested across the same set of switching events. It can be easily seen that the DRL-based method (blue dots) are much more closely clustered than the default method (red dots), which are distinctly spread out and have a distinct relationship whereby performance decreases (with respect to both mean and standard deviation) as the temperature gets colder. In particular, the performance degrades considerably for the default method at $5\degree C$. The middle image compares the Jensen-Shannon Distance (JSD) between the distribution of ST values measured at room temperature ($25\degree C$) and each other tested temperature. The JSD is a measure of similarity between distributions, where a smaller value (distance) implies more similar distributions; 0 indicates identical distributions and 1 indicates no statistical similarity between them. Ideally, the JSD between ST distributions at different temperatures will be as low as possible, since this would indicate that temperature variation does not move the performance very far away from it's room temperature performance. As the middle image in Figure \ref{fig:jensen_shannon} shows, the JSD of the DRL-based method is consistently lower for all temperatures than the default. Moreover, it also varies less dramatically over each temperature; the JSD varies between 0.4-0.45 for the DRL-based method vs 0.425-0.6 for the default ($3.5\times$ improvement). Finally, the right-most image in Figure \ref{fig:jensen_shannon} shows the mean (standard-deviation shown by the error-bars) difference between the $25\degree C$ test and each other temperature on a per-event basis. Perfect similarity is 0 mean with 0 standard-deviation (indicated by the dotted black line) and values are normalised to the $25\degree C$ measurements. It shows that the performance of each switching event does not change significantly on average across the variety of temperatures tested, with only a small penalty on standard deviation of per-event switching performance comparison on the two middle temperatures ($35\degree C\ \&\ 53\degree C$) on account of a small number of outlier events. For $35\degree$ and $53\degree$ the DRL method has almost an identical mean performance to the room temperature results, and a very small divergence at the largest temperature. By comparison the default method is always more divergent than the DRL method, and increasingly so the further away in temperature it is. This is a clear indication that under thermal variation, the DRL-generated PID parameters produce performance that is superior to the default method both with respect to the absolute performance, as well as self-consistency.

\section{Conclusion}

This paper has detailed and demonstrated a novel DRL-based methodology for offline one-shot multi-objective PID tuning. Demonstrated in the context of a real-world manufacturing environment operated by a market-leader in optical switching devices, the method achieves significant improvements of generating better control parameters with respect to the performance and thermal stability of resulting control processes. The method also generates control parameters $10^6\times$  more quickly than conventional methods. Finally, the method generates control parameters directly from generic physical information about device and can do so offline. This enables it to control each device uniquely with respect to it's distinct characteristics, as well as be implemented in parallel/in the background to other processes in production effectively eliminating optimisation overhead.

The method was demonstrated experimentally on a set of piezoelectric actuators in a large $384\times384$ optical switch which require fast and stable control process to ensure reliable communications. Compared to traditional tuning techniques, the proposed method achieves a $5\times$ improvement in the number of switching events that settle in less than or equal to the most challenging target time, as well as improving the mean and standard-deviation of the switching speed by 23\% and 43\% respectively. This is also achieved without compromising overshoot beyond what is operationally acceptable. Moreover the parameters generated by the proposed method achieve significantly more consistent performance compared to room temperature ($25\degree C$) when temperature varies between $5\degree C$ and $73\degree C$ in contrast to traditional tuning methods. Finally, with respect to implementation viability in a production context, the proposed method takes $\mathcal{O}(hour)$ to train and $\mathcal{O}(ms)$ to generate unique PID parameters for a device unseen during training. Traditional tuning methods can take $\mathcal{O}(day)-\mathcal{O}(week)$ to generate a single set of generic control parameters that must be applied to any other actuator unless the process is to be repeated. As such, the propsoed method achieves a $10^6\times$ speedup compared to industry-standard tuning methods. Moreover the method generates parameters using an input of only generic physical information about the device (resonance frequency, gain) which is reasonably assumed to be known after standard post-manufacturing characterisation/quality-checking processes. For this reason, devices do not have to be present nor operated by the method in order for optimal control parameters to be generated for them. As such, optimisation-overhead in the production process is reduced to virtually zero.

\section{Acknowledgements}

This work was supported under the Engineering and Physical Sciences Research Council (EP/R041792/1 and EP/L015455/1), the Industrial Cooperative Awards in Science and Technology (EP/R513143/1), the OptoCloud (EP/T026081/1), and the TRANSNET (EP/R035342/1) grants, as well as by Huber+Suhner Polatis Limited.

 \bibliographystyle{elsarticle-num} 
 \bibliography{bibliography}





\end{document}